\documentclass[prd, preprint, nofootinbib, groupedaddress]{revtex4}	

\usepackage{graphicx}
\usepackage{color}
\usepackage{acronym}
\usepackage[usenames,dvipsnames]{xcolor}
\usepackage[colorlinks=false,hyperfootnotes=false]{hyperref}
\usepackage{amsmath}
\usepackage[normalem]{ulem}

\newcommand{\unit}[1]{\ensuremath{\,#1}}
\newcommand{\unittext}[1]{\ensuremath{\unit{\mathrm{#1}}}}
\newcommand{\sqdeg}{\unittext{deg}^2}

\newcommand{\Lten}{\unit{L}_{10}}
\newcommand{\Lstar}{L^*}
\newcommand{\Mpc}{\unittext{Mpc}}
\newcommand{\cubicMpc}{\Mpc^{3}}
\newcommand{\percubicMpc}{\Mpc^{-3}}
\newcommand{\numpoints}{N}

\newcommand{\numtiles}{\mathcal{N}}

\newcommand{\successF}{\mathcal{F}}

\newcommand{\followup}{follow-up}

\makeatletter
\newcommand*{\defeq}{\mathrel{\rlap{%
                     \raisebox{0.3ex}{$\m@th\cdot$}}%
                     \raisebox{-0.3ex}{$\m@th\cdot$}}%
                     =}
\makeatother

\acrodef{FOV}{field-of-view}
\acrodef{GRB}{$\gamma$-ray burst}
\acrodef{GW}{gravitational-wave}
\acrodef{EM}{electromagnetic}
\acrodef{SNR}{signal-to-noise ratio}
\acrodef{GWGC}{Gravitational Wave Galaxy Catalog}
\acrodef{BNS}{binary neutron star}
\acrodef{CBC}{compact binary coalescence}

\begin{document}

\title{Utility of galaxy catalogs for following up gravitational waves from
binary neutron star mergers with wide-field telescopes}

\author{Chad Hanna}
\email{channa@perimeterinstitute.ca}
\affiliation{Perimeter Institute for Theoretical Physics, Waterloo, Ontario, N2L 2Y5, Canada}

\author{Ilya Mandel}
\email{imandel@star.sr.bham.ac.uk}

\author{Will Vousden}
\email{will@star.sr.bham.ac.uk}
\affiliation{University of Birmingham, Edgbaston, Birmingham, B15 2TT, United Kingdom}

\begin{abstract}
    The first detections of gravitational waves from binary neutron star mergers with advanced LIGO and Virgo observatories are anticipated in the next five years.  These detections could pave the way for multi-messenger gravitational-wave (GW) and electromagnetic (EM) astronomy if GW triggers are successfully followed up with targeted EM observations.  However, GW sky localization is relatively poor, with expected localization areas of $\sim 10$--$100\sqdeg$; this presents a challenge for following up GW signals from compact binary mergers.  Even for wide-field instruments, tens or hundreds of pointings may be required.  Prioritizing pointings based on the relative probability of successful imaging is important since it may not be possible to tile the entire gravitational-wave localization region in a timely fashion.  Galaxy catalogs were effective at narrowing down regions of the sky to search in initial attempts at joint GW/EM observations.  The relatively limited range of initial GW instruments meant that few galaxies were present per pointing and galaxy catalogs were complete within the search volume.  The next generation of GW detectors will have a ten-fold increase in range thereby increasing the expected number of galaxies per unit solid angle by a factor of $\sim 1000$.  As an additional complication, catalogs will be highly incomplete.  Nevertheless, galaxy catalogs can still play an important role in prioritizing pointings for the next era of gravitational-wave searches.  We show how to quantify the advantages of using galaxy catalogs to prioritize wide-field follow-ups as a function of only two parameters: the three-dimensional volume within the field of view of a telescope after accounting for the GW distance measurement uncertainty, and the fraction of the GW sky localization uncertainty region that can be covered with telescope pointings.  We find that the use of galaxy catalogs can improve success rates by $\sim 10\%$ to a factor of $4$ relative to follow-up strategies that do not utilize such catalogs for the scenarios we considered.  We determine that catalogs with a $75\%$ completeness perform comparably to complete catalogs in most cases, while $33\%$-complete catalogs can lead to lower follow-up success rates than complete catalogs for small fields of view, though still providing an advantage over strategies that do not use a catalog at all.
\end{abstract}

\maketitle

\section{Introduction\label{sec:introduction}}

\citet{S6BurstFollowup, lowlatency} present the first {\em low-latency} searches for gravitational
waves that triggered \ac{EM} \followup{} observations with a $\sim 30$-minute
response time\footnote{
$\lesssim 5\unittext{minutes}$ not including a human intervention time scale
that could be removed if automated.
}.  No gravitational waves were detected, but \ac{GW} candidate events
consistent with noise were followed up successfully \cite{SwiftS6, LVCOpticalTransients}.  
Later this decade, a
network of advanced \ac{GW} detectors including LIGO and Virgo \cite{AdvLIGO,
AdvVirgo} may detect tens of \ac{BNS} mergers per year once at full sensitivity
(with a plausible range of one detection in a few years to a few hundred
detections per year) \citep{ratesdoc}.  Some of these detections may be
accompanied by \ac{EM} counterparts \citep[e.g.,][]{Metzger2012, Kelley:2012,
Bloom:2009}, summarized below.  Several nearly ``instantaneous'' search methods
for \acp{GW} from \ac{BNS} mergers have been proposed, introducing the
possibility of transmitting information about the candidate to \ac{EM} telescope
partners within tens of seconds of a binary merger \citep{LLOID, Luan2012}.

%
\ac{BNS} mergers are thought to generate several distinct \ac{EM} counterparts
spanning most of the \ac{EM} spectrum; Figure 1 of \citep{Metzger2012} illustrates the counterpart
emission mechanisms.  Short, hard gamma ray bursts occur on timescales of $\lesssim 2$ seconds \citep{Nakar:2006} and are strongly beamed.  Afterglows from shock waves produced when the emitted jet encounters the interstellar medium span the spectrum from X-rays to radio waves \cite[e.g.,][]{vanEertenMacFadyen:2011, Berger:2010, Perley2009, NakarPiran:2011}.   Thermal emission from $r$-process nucleosynthesis in the merger ejecta has been predicted to peak in the 
infrared \cite{Kasen2013}; the first hint of such a kilonova signal has been recently observed \citep{Tanvir2013}.

Several transient telescope networks exist with wide-field coverage and it is
important to understand what is the best way to tile pointings within the
\ac{GW} localization region using wide-field instruments.  This question has
been addressed partly by \citet{Singer2012}, who present a framework for
allocating telescope resources to optimally cover the available sky localization
region.  In this work we consider the situation in which only a fraction of this
area can be surveyed in a timely fashion, where it is important to choose the
tiles that represent the most likely source location first.  Both
\citet{Singer2012} and \citet{Fairhurst:2009} focus on the assumption of a
uniform-on-the-celestial-sphere prior on the \ac{GW} source location.  However,
given the broad \ac{GW} localization region, pointing might be strongly
influenced by a sharply peaked prior expectation for the signal location.
\citet{Fairhurst:2009} mentions that a galaxy catalog could serve as a better
prior and, indeed, \citet{S6BurstFollowup, lowlatency} demonstrate that the use of a galaxy
catalog \citep{LIGOS3S4Galaxies} greatly increases the chance of imaging an
\ac{EM} counterpart in simulations for the initial \ac{GW} detectors with
$\lesssim 20\Mpc$ range for \acp{GW} from merging \acp{BNS}.  At this range,
nature provides few galaxies as potential hosts for the merger, corresponding to
sharp peaks in the prior probability.

The same angular scale will encompass many more galaxies in the advanced GW
detector era and the usefulness of a galaxy catalog prior comes into question.
\citet{Metzger2012} suggest that the number of bright galaxies in the
localization region will be too large to improve the prospects of imaging the
\ac{EM} counterpart.  For example, \ac{GW} detections in the advanced
detector era will occur at a median distance of $\sim200\Mpc$.  A source at this distance may be optimistically localized to a sky area of $20\sqdeg$ and a fractional distance error of $\sim 30\%$ by \ac{GW} measurements alone~\citep{Fairhurst:2009, scenarios,
Veitch:2012, Nissanke:2011, Rodriguez2013} with a network of three or more \acp{GW} detectors\footnote{With significantly larger uncertainties expected for a two-detector network for the early runs of Advanced LIGO alone \cite{scenarios, KasliwalNissanke:2013}.}.  The volume defined by this solid angle and distance range will contain more than $500$ galaxies
brighter than $0.1\Lstar$ (see Section~\ref{sec:tilefluctuations}) -- more than
can realistically be imaged individually on short timescales. 

If wide-field instruments are used to tile the \ac{GW} localization region and
the requirements on the speed and depth of the search make it impossible to follow up the entire
localization region, the question arises of how to prioritize which tiles should be observed.
\citet{NuttallSutton:2010} partly address this problem by simulating follow-up
searches within $100\Mpc$ in the advanced detector era using the \ac{GWGC} of
\citet{White2011}.  Individual galaxies are targeted on the basis of a ranking
algorithm that accounts for luminosity and distance to putative host galaxies.
Meanwhile, \citet{Nissanke2013} provide case studies for the process of
detecting a \ac{GW} event and locating and identifying its \ac{EM} counterpart,
using galaxy catalogs to eliminate false-positive \ac{EM} signals.  Both studies
find that catalogs can be useful both for locating and identifying an
\ac{EM} counterpart when there are insufficient resources to point individually
at each galaxy.


In this work, we revisit the utility of a galaxy catalog in the regime where
there are too many galaxies in the \ac{GW} localization region to be followed up
individually, and observational constraints on the speed and depth of the search 
prevent complete coverage with
wide-field instrument pointings.  We quantify the utility of a galaxy catalog as a function of 
the three-dimensional volume within the \ac{FOV} of the follow-up telescope (after accounting for the distance measurement uncertainty from \ac{GW} measurements) and of the fraction of the \ac{GW} localization region that can be covered.  We consider realistic catalogs, which are likely to be significantly incomplete within the large volumes in which advanced detectors are sensitive.

We find that even in the advanced-detector
era, galaxy catalogs can still confer benefits through the inherent fluctuations
in luminosity density on the sky.  Galaxy luminosity and count fluctuations will
help to prioritize tiles and increase the relative probability of imaging a
\ac{GW} \ac{EM} counterpart.  In particular, we will show that catalogs are most
relevant for narrow and shallow follow-up searches (that is, smaller \acp{FOV}
and shorter range) and that improving the completeness and range of existing
catalogs is important for \ac{EM} follow-up efforts.

This work is organized as follows.  In Section~\ref{sec:problem-statement} we
define our condition for a ``successful'' follow-up and describe the algorithm
we use to select tiles for pointing, given a galaxy catalog.  In
Section~\ref{sec:tilefluctuations} we discuss the characteristics of the
galaxy luminosity distribution and show that there can be significant variations
in luminosity between tiles.  Sections~\ref{sec:results-complete} and \ref{sec:incompleteness} present the results of simulated follow-up searches for several detection and observation
scenarios and discuss the effects of incompleteness of the galaxy catalog on the
results.  Section~\ref{sec:conclusion} concludes with a brief discussion of our results
and additional suggestions for future work.

\section{Problem statement\label{sec:problem-statement}}

In this work we are not concerned specifically with identifying host galaxies,
but rather with choosing the most probable sky regions commensurate with a given
\ac{FOV} by using galaxy catalog information.  We neglect many of the
practicalities considered by \citet{Nissanke2013} and \citet{Singer2012} (e.g.
telescope slew time, limiting depth, day/night observation time, etc.) to
isolate the utility of galaxy catalogs on their own merits.  We do, however,
assess the effect of incompleteness of galaxy catalogs in our method.

Throughout this work we will use a blue-band galaxy catalog as a proxy for merger rate density.
This assumes that the rate of \ac{BNS} mergers is proportional
to the instantaneous massive star formation rate (with negligible time delays between
formation and merger) and is therefore tracked by blue-light
luminosity~\cite{Phinney:1991ei}.  On the contrary, observational evidence indicates that a quarter of short gamma ray bursts occur in elliptical galaxies with no signs of ongoing star formation \cite{FongBerger:2013}.  However, the choice of color is not critical for the modeling below; it is sufficient to assume that we have a catalog that is an accurate tracer of merger rate density.  We discuss the validity of this assumption in Section~\ref{sec:conclusion}.

To model the effect of using galaxy catalogs to assist in EM followup we begin
by dividing the GW localization area, $A$, into $\numtiles$ tiles (assumed to be
non-overlapping for simplicity), each representing a telescope \ac{FOV} $P$,
where $\numtiles=\lceil\tfrac{A}{P}\rceil$.

We define a {\bf successful \followup{}} as a \acs{GW}-triggered \ac{EM}
transient search in which one of the tiles selected for imaging
contains the \ac{GW} source.  For simplicity, we require only that the source
{\em reside} in one of the tiles, and not that the expected \ac{EM} counterpart
is actually {\em detectable} by a given \followup{} instrument\footnote{
In fact, the depth to which the available telescopes can detect an \ac{EM}
transient may influence the optimal choice of \followup{} target.  For example,
there is little point in targeting galaxies that are so distant that the
transients they might contain would not be detectable by a given telescope.
} or distinguishable from background events.  We therefore assume the
transient search to be limited in range only by the capabilities of the \ac{GW}
detector network and not by the depth of the
\followup{} instrument.  
Considering the above assumptions, the probability of success
is $1$ if all tiles in the sky are searched, regardless of whether the
correct transient is identified.  

We define the success fraction $\successF$ as the fraction
of \ac{GW} events that are expected to be successfully followed up for a given follow-up strategy according to the definition above.  If one ignores the galaxy distribution in
the event localization area\footnote{
In practice, not all pointings in the sky will have the same likelihood; indeed,
in the high \ac{SNR} limit, the likelihood distribution will have a Gaussian
shape.  The probability density function on the sky will be computed through
coherent parameter estimation on \ac{GW} detector data \cite{S6PE}; here,
we treat $A$ as a suitable ``effective'' area.
} $A$, the relative probability that a \ac{GW} is in a
given tile is uniform amongst the tiles. The success fraction is
\begin{equation}
    \successF = \sum_i^{\numpoints} \frac{1}{\numtiles} = \frac{\numpoints}{\numtiles} \equiv f,
\end{equation}
where $\numpoints$ is the number of telescope pointings compatible with search speed and depth requirements, and $f$ is simply the fraction of the GW localization area that is followed up,
$f=\numpoints\tfrac{P}{A}$.  

This should be compared to the case where each tile has
a relative probability of containing the \ac{GW} event proportional to its blue
light luminosity $L_i$, and a greedy pointing algorithm is used whereby the
brightest tiles are pointed at first, $L_i \ge L_{i+1}$ for all $i$:
\begin{equation}
    \successF = \frac{1}{L} \sum_i^{\numpoints} L_i \, ,
\end{equation}
where $L \equiv {\sum_i^{\numtiles} L_i}$.  With the greedy strategy, $\successF \geq
f$;  in other words, if \ac{GW} sources are distributed according to blue
luminosity then using that information never hurts the success fraction.

The \ac{GW} amplitude depends on the inclination and orientation of the source relative to the line of sight, with the highest detector response for face-on sources.  This allows us to compute the probability that a source in a given galaxy at a known distance and sky location would pass a signal-to-noise-ratio detection threshold under the assumption that the binary's inclination and orientation are isotropically distributed.  This probability decreases from $\approx 1$ for a very nearby galaxy to $0$ for a galaxy at the maximum distance for a given sky location and detector network configuration (where only face-on sources would be detectable); the decrease is roughly linear in the distance to the source (cf.~the ad hoc \citep{NuttallSutton:2010} weighting of galaxies in the catalog by one over distance or one over distance squared).
In principle, this detection probability should be included in the prior weighting of galaxies in the catalog, giving each galaxy an effective luminosity that is the product of its actual luminosity and the probability that a source in this galaxy would be detectable in a \ac{GW} search with the given detector network.  

However, in practice, the analysis of \ac{GW} data will yield (strongly correlated) constraints on  distance and inclination, so this prior probability should not be assigned independently of the detector data.  As discussed in Section~\ref{sec:conclusion}, the correct approach would be to include the galaxy catalog directly in coherent Bayesian parameter estimation as a prior, which would allow for a self-consistent application of all information, rather than attempting an a posteriori correction as we are doing here.  However, for the purposes of estimating the utility of a galaxy catalog, we take the simplified approach of considering only galaxies in a range of distances consistent with the distance measurement accuracy expected for multi-detector networks: $\sim 30\%$ in fractional distance uncertainty for an event at the detection threshold \cite{Veitch:2012}.  Within this range, we will neglect the detection probability in the galaxy prior, and consider only priors proportional to blue-light luminosity.  We expect this to be conservative, since the effective galaxy luminosity with the detection probability included would have had greater fluctuations than the absolute luminosity, and, as we will see below, luminosity fluctuations increase the utility of galaxy catalogs.

We will thus assume that the detector network is able to localize a source at distance $D$ to within a range $\in [D_\textrm{min}\, , D_\textrm{max}]$, with $D_\textrm{min}=0.7 D$ and $D_\textrm{max}=1.3 D$.  Combining the solid angle $P$ of a telescope pointing with this range, we can define the {\em pointing volume}, i.e., the volume of each pointing within the measured distance range, as $V \equiv
\tfrac{4}{3}\pi (D_\textrm{max}^3-D_\textrm{min}^3)\,\tfrac{P}{\Omega}$ where $\Omega\approx41,000\sqdeg$ is the
solid angle of the whole sky. The average
luminosity per pointing volume is then given by $\langle L_i \rangle = V
\rho_L$, where $\rho_L$ is the average spatial density of luminosity.  We will use a luminosity
density $\rho_L = 0.02 \Lten\percubicMpc$, where $\Lten$ is defined as $10^{10}$
times the solar blue-light luminosity $\unit{L_{B,\odot}}$
\cite{LIGOS3S4Galaxies, ratesdoc}.

\section{Luminosity fluctuations\label{sec:tilefluctuations}}
 
\begin{figure}[htb]
    \includegraphics{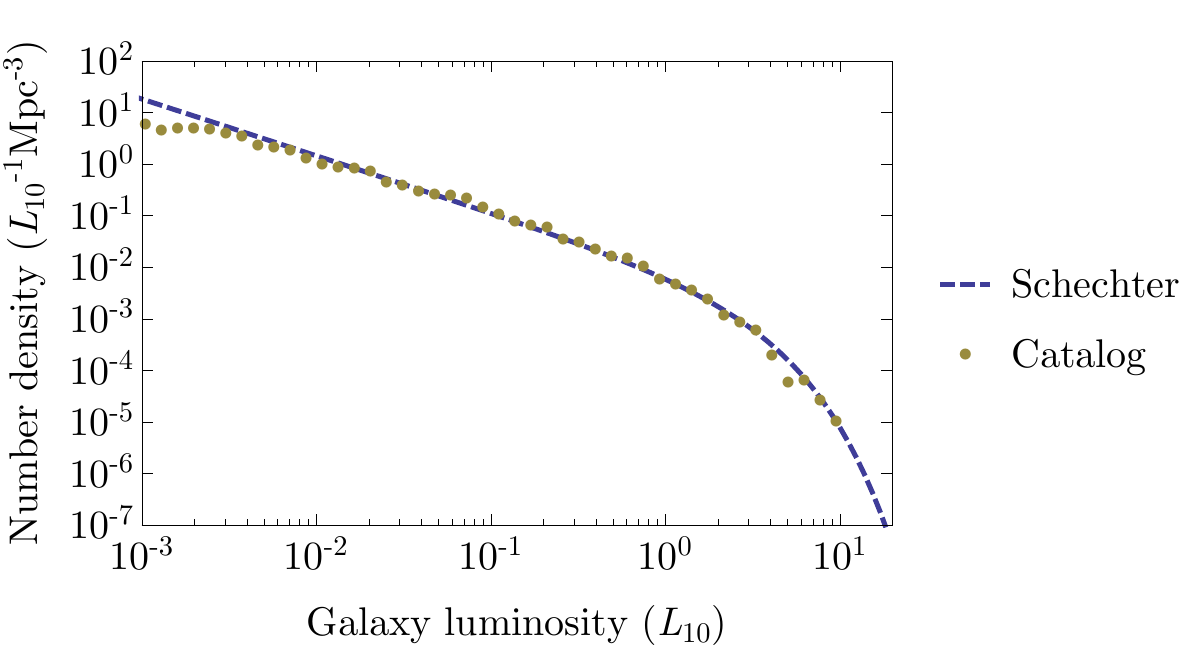}
    \caption{\label{fig:lumfunc} GWGC catalog luminosity function within
        $20\Mpc$ compared to a fit of \eqref{eq:lumfunc} with $\Lstar=2.2\Lten$
        and $\alpha=-1.1$.}
\end{figure}

In this section, we incorporate the distribution of intrinsic galaxy luminosity and the counting fluctuations in the number of galaxies in different pointings into the expected distribution of $L_i$.
We neglect spatial correlations of galaxies (e.g., due to the presence of galaxy
clusters -- a conservative assumption since greater clustering improves the
utility of galaxy catalogs, as we shall see shortly) and assume they are homogeneously distributed in volume.  We model the distribution of galaxies in blue luminosity and volume as a Schechter function~\citep{Schechter}
\begin{equation}
    \label{eq:lumfunc}
    n(x)\,dx \propto x^\alpha e^{-x} \,dx \, ,
\end{equation}
where $x \equiv L/\Lstar$ and $n(x)\,dx$ is the expected number of galaxies per
$\cubicMpc$ in the interval $[x,x+dx]$.  We use the \ac{GWGC}~\cite{White2011}
within $20\Mpc$, where it is complete, to estimate $\alpha=-1.1$ and $\Lstar=2.2 \Lten$, slightly
brighter than the Milky Way's blue-band luminosity of $\sim1.7 \Lten$.\footnote{Similar values of these parameters are quoted in the literature, e.g., $\alpha=-1.07$, $\Lstar=2.4 \Lten$ \cite{Schneider:2006}; our conclusions are insensitive to small changes in these parameters.}
We
normalize the luminosity function to yield $\rho_L = 0.02 \Lten\percubicMpc$ on
the interval $L \in [0.001, 20] \Lten$ \citep{Kopparapu2008}.  All following results are based on this Schechter luminosity distribution.  Figure~\ref{fig:lumfunc} shows the luminosity function of the \ac{GWGC} within $20\Mpc$ as well as the Schechter model.

\begin{figure}[htb]
    \includegraphics{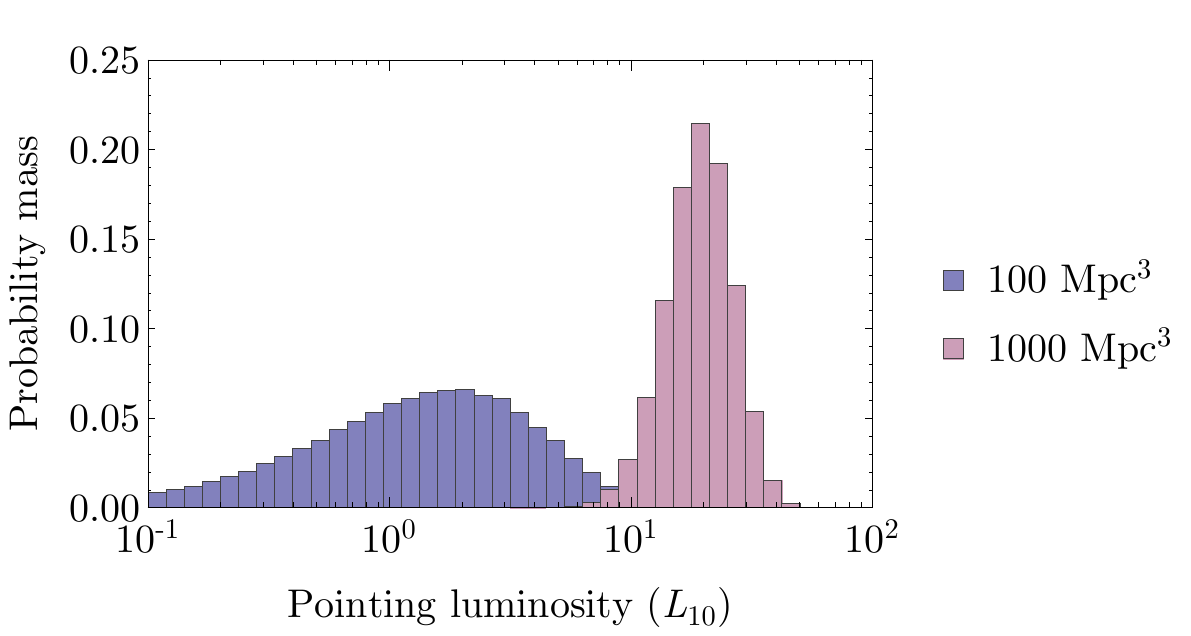}
    \caption{\label{fig:Lhist} Distribution of luminosity drawn from
        \eqref{eq:lumfunc} for fixed volumes of $100\cubicMpc$ and
        $1000\cubicMpc$.  The means are $2.0\Lten$ and $20.0\Lten$ and the
        standard deviations are $2.0\Lten$ and $6.3\Lten$ respectively.  These
        values correspond approximately to a Poisson scattering of galaxies of
        ``typical'' luminosity $\Lstar$.}
\end{figure}

If we assume that a pointing tile has volume $V$, containing a random integer sample of
galaxies taken from the distribution in \eqref{eq:lumfunc}, then the resulting
luminosity $L$ in that volume can be described by a random variable of mean
$V\rho_L$.  The results of a direct Monte Carlo simulation of \eqref{eq:lumfunc} for
$100\cubicMpc$ and $1000\cubicMpc$ are shown in Figure~\ref{fig:Lhist}.  To understand these results, one can crudely approximate the Schechter galaxy population as a
Poisson scattering of identical galaxies of ``typical'' luminosity $\Lstar$.  In
this case, the luminosity in a volume $V$ is simply $L=n\Lstar$, where $n$ is
drawn from a Poisson distribution of mean $V\rho_L/\Lstar$.  For $100\cubicMpc$
and $1000\cubicMpc$ pointing volumes, for example, we should expect $0.9 \pm 0.95$ and $9\pm 3$ galaxies per pointing volume, with corresponding luminosities of $2.0\pm2.1\Lten$ and $20\pm6.6\Lten$, respectively.  This closely matches the fluctuations of $2.0\pm2.0\Lten$ and $20\pm6.3\Lten$, respectively, measured via a Monte Carlo simulation of the actual Schechter distribution.    

The large variations in tile luminosity -- $2\pm2\Lten$ for the $100\cubicMpc$
volume -- suggest that there can be substantial advantage to following up the
brightest tiles first in a survey with limited pointings.  For lower
pointing volumes, the distribution becomes increasingly non-Gaussian, and its
skewness amplifies the advantage of luminosity-directed surveys.

\section{Results: complete galaxy catalog\label{sec:results-complete}}

\begin{figure}[htb]
    \includegraphics{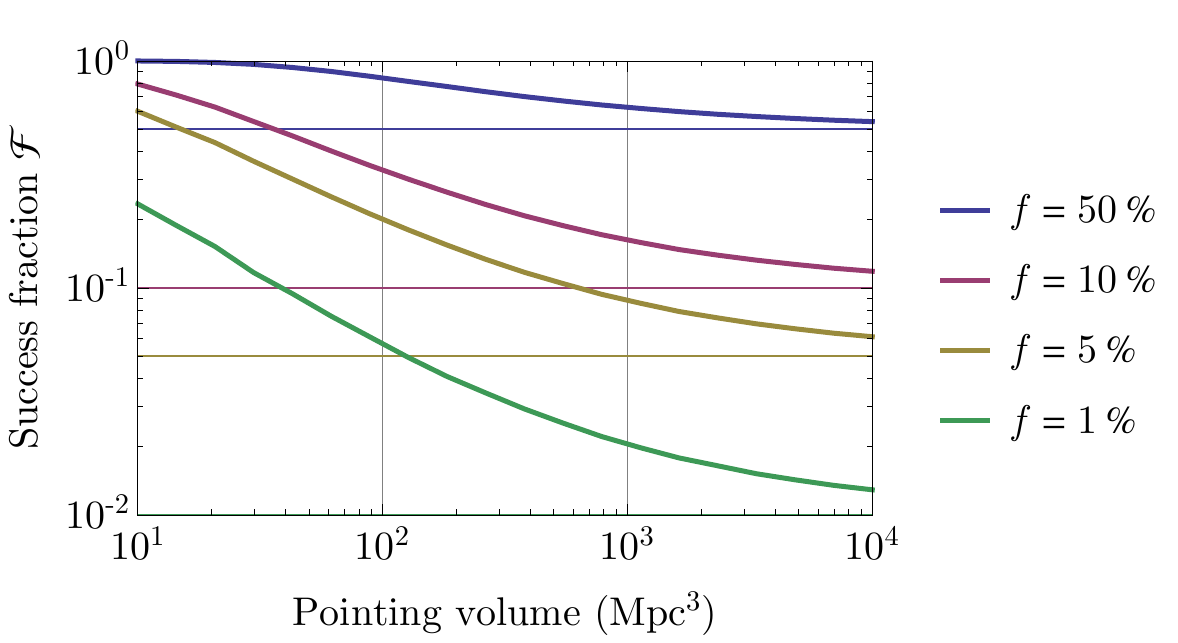}
    \caption{\label{fig:FvsV} A comparison of the success fraction $\successF$
        relative to the follow-up fraction as a function of pointing volume.  
        The horizontal grid lines represent
        follow-up searches that do not use galaxy catalogs, wherein
        $\successF=f$.  The large variation in
        luminosity per tile can cause certain \ac{FOV}s within the GW localization
        region to be more likely to contain the source, suggesting an obvious pointing
        priority in the case where the entire localization region cannot be
        followed up in a timely fashion.  }
\end{figure}

We show the success fraction $\successF$ when using a complete, ideal galaxy catalog as a function of pointing volume in Figure~\ref{fig:FvsV}, for four
choices of the fraction $f \in \{0.01, 0.05, 0.10, 0.50\}$ of the \ac{GW}
localization region being followed up.
Recall from Section~\ref{sec:tilefluctuations} that in
the case where no galaxy catalog is used we would expect on average that
$\successF = f$.  We find in Figure~\ref{fig:FvsV} that in all cases when using
the galaxy catalog $\successF > f$ as we would expect, with the advantage of the catalog being
more pronounced for smaller pointing volumes where the variation of luminosity
per pointing is larger.

It is useful to apply the results of Figure~\ref{fig:FvsV} to a few potential
scenarios in order to understand the impact that an ideal galaxy catalog would
have.  The distance at which a single advanced LIGO detector is capable of detecting an
optimally oriented and located \ac{BNS} merger at an \ac{SNR} of $8$ -- known as
the {\em horizon distance} -- is $\sim 450\Mpc$ \citep{ratesdoc}.  However,
averaging over sky locations and orientations, we expect $75\%$ of detections to
come from within $\sim250\Mpc$, or $50\%$ from within $\sim200\Mpc$.  For early
versions of the advanced LIGO/Virgo network, this could be reduced to as little
as $D\sim100\Mpc$ for the $50\%$ percentile \citep{scenarios}.   We therefore consider the median distance of $200$ Mpc as a typical distance to a detection with the advanced detector network
operating at design sensitivity in cases~\ref{item:case-1} and \ref{item:case-2}.  Meanwhile,  
case~\ref{item:case-3} represents the rarer scenario of a closer
source at an estimated distance of $100$ Mpc.  

\begin{enumerate}
    \item Consider a $10\sqdeg$ \ac{FOV} telescope following
        up with a single pointing a \ac{GW} source estimated to be at a distance of $200\Mpc$, with
        a $100\sqdeg$ localization region.  The pointing volume to this source, assuming \ac{GW} observations constrain the distance to be $\in[140,260]$ Mpc, is $15,000\cubicMpc$.  (For comparison, a $1 \sqdeg$ conical \ac{FOV} contains a volume of $\sim 100\cubicMpc$ out to a distance of $100\Mpc$.)  
        Without a galaxy catalog we
        would expect that the success fraction $\successF = f = 10\%$.  However,
        Figure~\ref{fig:FvsV} shows that using a galaxy catalog we might expect
        to have a $\sim 11\%$ success fraction.
        \label{item:case-1}
    \item Consider the same \ac{GW} source as case 2 but with a $1\sqdeg$
        \ac{FOV} follow-up instrument having 10 pointings.  While the overall
        coverage is still $f=10\%$, the pointing volume is reduced to $1500\cubicMpc$, so the
        fluctuations in luminosity between tiles are more significant. As a
        result, the success fraction improves to $\sim 16\%$.
        \label{item:case-2}
    \item Finally, consider a loud \ac{GW} signal with a $100\Mpc$ distance estimate being
        followed up by a single pointing of a $1\sqdeg$ \ac{FOV} instrument.  The 
    sky-localization and distance measurement accuracy improve for high signal-to-noise ratio \ac{GW} detections.  We therefore consider a $10\sqdeg$ localization region and a reduced distance uncertainty range $\in[90,110]$ Mpc.  In this case, the pointing volume is only $60\cubicMpc$, and the success fraction is $\sim 40\%$, a
        $4$-fold improvement over the nominal $\successF=f=10\%$ success
        fraction in the absence of a galaxy catalog.
        \label{item:case-3}
\end{enumerate}

These cases are meant as illustrations only.  Distances to optimally located and oriented sources may range to $450$ Mpc for advanced detectors at design sensitivity.  Meanwhile, source in the early phases of advanced detector commissioning, when detectors are sensitive within a smaller range, may resemble case \ref{item:case-3} in typical distance estimates, but with poorer sky localization and distance measurements.  

While the overall fraction $f$ of the \ac{GW} localization region is $10\%$ for
each of the above cases, the pointing volumes are respectively $\sim
15,000\cubicMpc$, $\sim 1500\cubicMpc$, and $\sim 60\cubicMpc$.  The progressively
larger success fractions for each case illustrates how the utility
of the catalog depends on pointing volume.

The effectiveness of a given follow-up telescope, as characterized by its
\ac{FOV} $P$ and the number of pointings $N$ that can be taken within the allotted time while observing to a sufficient depth, may be influenced by the sensitivity of 
the \ac{GW} search and the distance estimate it yields.  
For case~\ref{item:case-3}, an instrument with a larger
\ac{FOV} might be chosen at the expense of depth, since the \ac{EM} signal is
expected to be louder.  Similarly, the greater imaging depth required to detect
transients at $200\Mpc$ might mean that fewer pointings are available for the
more distant sources in cases~\ref{item:case-1} and \ref{item:case-2}.

\section{The effect of galaxy catalog completeness\label{sec:incompleteness}}

\begin{figure}[htb]
    \includegraphics{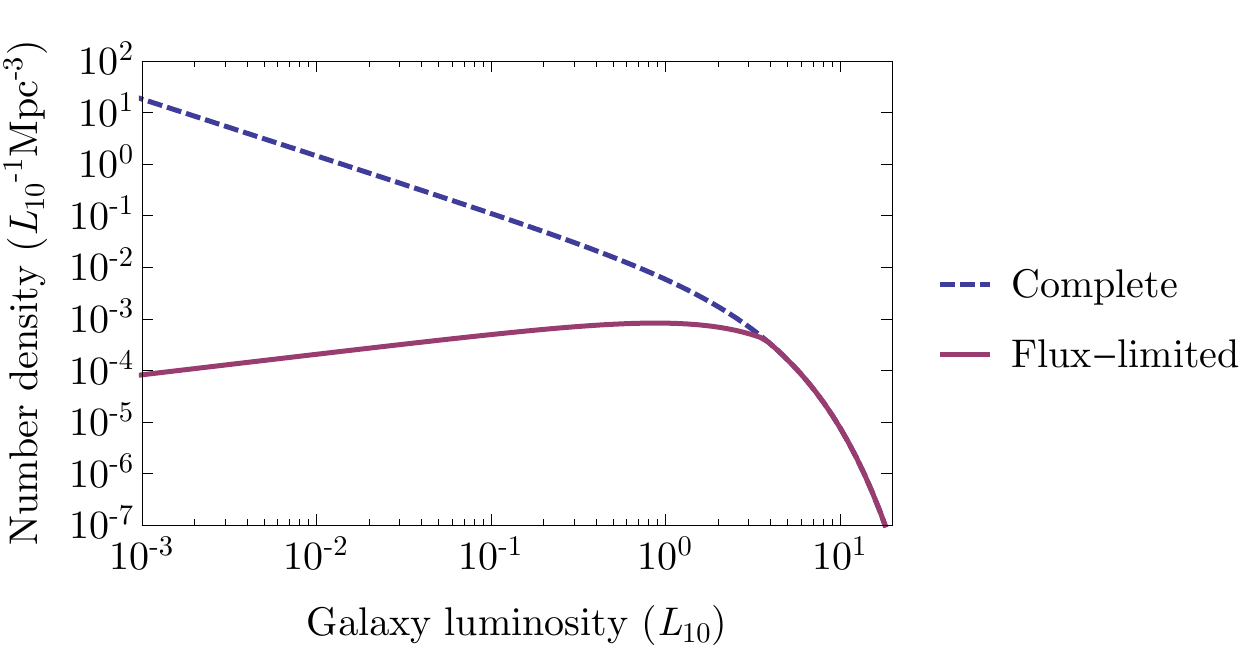}
    \caption{\label{fig:inlumfunc} Comparison of an ideal complete luminosity
        function (dashed) to a hypothetical flux-limited survey (solid) at
        apparent magnitude $m_B=15.5$ out to $200\Mpc$.  This example is only
        $33\%$ complete relative to the expected $0.02 \Lten\percubicMpc$.}
        %
	%
\end{figure}

The previous discussion assumed that an ideal, complete galaxy catalog was
available; however, \ac{GWGC} \cite{White2011},
is incomplete beyond $\sim 30\Mpc$, and there are limitations to how complete
catalogs become at $\sim 200 \Mpc$ distances \cite{Metzger2013}.  
In practice, catalogs may comprise many different surveys with different characteristics and selection criteria, and will be influenced by spatially dependent factors such as extinction in the
Galactic plane.  However, the simplest model, and the one we consider here, is
incompleteness from a flux-limited survey.  As a simple example, we considered
an extremely flux-limited survey that does not resolve galaxies fainter than
apparent magnitude $\sim 15.5$ in the blue band, which is a rough approximation
to the \ac{GWGC}.  The resulting luminosity function of this hypothetical
survey, for galaxies within $200\Mpc$, is shown in Figure~\ref{fig:inlumfunc}.
The catalog is only $33\%$ complete within this volume (i.e., contains $33\%$ of
the total absolute luminosity) compared to the model presented in
Figure~\ref{fig:lumfunc}.  However, it contains most of the rare bright
galaxies, whose distribution on the sky shows significant fluctuations, making them most useful for informing pointing strategy, while missing
common dim galaxies which are nearly homogenous on the sky and are therefore less useful for pointing.

\begin{figure}[htb]
    \includegraphics{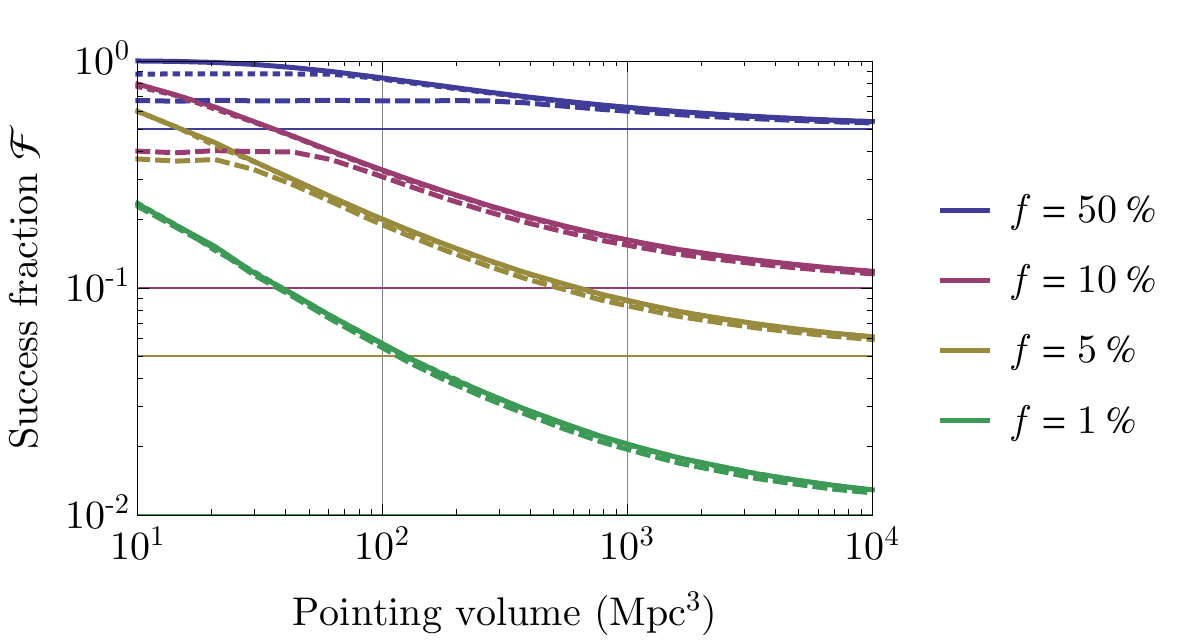}
    \caption{\label{fig:inFvsV}The success fraction $\successF$ as a function
        of pointing volume for hypothetical flux-limited catalogs of $33\%$
        (dashed), $75\%$ (dotted) and $100\%$ (solid) completeness within the
        pointing volume being considered.  Incomplete catalogs yield similar success
        fractions to complete catalogs except at small pointing volumes and
        large follow-up fractions.}
\end{figure}

The shape of the luminosity function for a flux-limited catalog is sensitive
only to the overall completeness of the catalog, not the specific range and
cut-off magnitude.  Therefore, in order to express the follow-up success
probability for a flux-limited catalog in terms of pointing volume, which
incorporates \ac{FOV} and depth in a single variable, we fix the {\em
completeness} of the catalog, rather than the cut-off magnitude.  This success
probability is plotted in Figure~\ref{fig:inFvsV} for three choices of
completeness: $33\%$, $75\%$, and $100\%$.  

In our flux-limited survey model for incompleteness, the catalog luminosity
function agrees with a hypothetical complete luminosity function for the most
luminous galaxies.   It is therefore not surprising that incompleteness in a
catalog has little effect on the scenarios where only a small fraction of the
sky uncertainty region will be followed up, since both complete and incomplete catalogs will tend
to agree on the most luminous tiles, which are the only ones that will be
pointed at for small follow-up fractions.  For example, in
Figure~\ref{fig:inFvsV}, the line corresponding to a follow-up fraction of
$f=0.01$ is virtually unchanged from the corresponding line for a complete
catalog.

When the follow-up fraction is large, incomplete catalogs still yield similar
success fractions to complete catalogs as long as the pointing volume is also
sufficiently large.  Of course, at very large pointing volumes $\successF$
asymptotes to $f$, as the fields of view become increasingly uniform due to the
very large number of galaxies they contain, and a galaxy catalog ceases to be
useful even when complete.  Even at moderate pointing volumes and moderate
follow-up fractions, catalog incompleteness is not necessarily a concern if it
only leads to missing the many dim galaxies which are nearly homogeneously
distributed on the sky.  A few bright galaxies can still dominate the prior and
since these are included even in incomplete flux-limited catalogs, the success
fraction is still relatively insensitive to completeness.  When pointing volumes
are small, even the dimmest galaxies, which are missed out in incomplete
catalogs, contribute to the variability between different fields of view.  When
follow-up fractions are large at small pointing volumes, the success rate
asymptotes to the maximum possible success fraction $\successF \to \lambda + f
(1-\lambda)$, where $\lambda$ is the catalog completeness\footnote{
Consider a situation in which the number of bright galaxies that enter the
catalog is sufficiently small that all of them can be followed up with a
negligible number of pointings; the remaining allowed pointings will capture a
fraction $f$ of the other fields of view, which have no galaxies in the
incomplete catalog and have a uniform prior density $1-\lambda$ painted across
them.
}, and incompleteness limits catalog utility.  This happens for a $33\%$-complete catalog when the pointing volume is $V \lesssim 100\cubicMpc$ and the follow-up fraction is $f\gtrsim 10\%$.  Nevertheless, even for larger follow-up fractions, $\successF$ is significantly
larger than $f$ for a large range of pointing volumes, suggesting that even a
moderately complete ($33\%$ complete) catalog is still useful for pointing at the
sky region hosting the source of a \ac{GW} transient.

As discussed above, an incomplete catalog is most useful when it contains a high
fraction of intrinsically luminous galaxies at the expense of missing galaxies
with low absolute magnitudes.  Therefore, a simple flux limit is an optimistic
model of a catalog's incompleteness.  If a catalog instead has a more gradual cut-off
with apparent magnitude, its utility for a given completeness fraction $\lambda$
could be lower than estimated here.  For example, for $f=10\%$, the largest discrepancy between a $33\%$-complete flux-limited catalog and GWGC is at $V\sim20\cubicMpc$, where they yield success fractions of $\successF\sim63\%$ and $\successF\sim60\%$, respectively.  The difference drops to $1\%$ for a pointing volume of $100\cubicMpc$ and is below the statistical error of the simulation for $V=1000\cubicMpc$.

\section{Conclusion and future work\label{sec:conclusion}}

Electromagnetic follow-up prospects in the advanced \ac{GW} detector era can be
aided by the use of galaxy catalogs to direct follow-up surveys.  The relevance of
catalog-directed wide-field follow-ups is limited mostly by the modest spatial
fluctuations of luminosity on the sky for the large three-dimensional localization uncertainty 
volumes of the advanced-detector network.

We have shown in Figures~\ref{fig:FvsV} and \ref{fig:inFvsV} that the utility of
a catalog depends on the volume of individual telescope pointings and on the
fractional coverage of the \ac{GW} localization area.  Catalogs are therefore
most relevant for shallow and narrow follow-up searches, although narrow-field
instruments are unlikely to follow up a sufficient fraction $f$ of the \ac{GW}
localization region for a successful follow-up to be realistic (with the
possible exception of short-range observations, where individual galaxies could
be followed up).  Loud, nearby \ac{GW} triggers are an obvious scenario where
catalogs will be particularly useful.  It is possible, for example, that they
confer as much as a four-fold increase in success fraction over a follow-up
that does not use a catalog; c.f. case~\ref{item:case-3} in
Section~\ref{sec:tilefluctuations}.  Similarly, follow-ups from shallower
\ac{GW} searches -- during the early commissioning phases of advanced detectors, for
example -- will also benefit from the use of catalogs.

However, even for sources located at the median $200\Mpc$ distance expected for detections with advanced-detector networks, we have shown that catalogs are still relevant for sufficiently small
telescope fields of view.  For example, a catalog might confer as much as a
$70\%$ increase in the probability of imaging the EM counterpart relative
to a follow-up without the benefit of a catalog, as in case~\ref{item:case-2}
of Section~\ref{sec:incompleteness}.

Realistic, incomplete galaxy catalogs are likely adequate for most follow-up
campaigns.  \citet{Metzger2013} propose that a catalog complete to $\sim75\%$
with respect to B-band luminosity should be achievable.  At $f=10\%$, a
hypothetical flux-limited catalog of this completeness concedes a fraction
$<1\%$ of the success fraction from a complete catalog for both $100\cubicMpc$
and $1000\cubicMpc$ pointing volumes.  \citet{Metzger2013} suggest that it will
be difficult to construct a galaxy of more than $\sim33\%$ completeness with
respect to K-band luminosity, a tracer of total mass.  Even in this case, the
fractional loss of success fraction relative to a search with a complete catalog
is small: $7.5\%$ and $5\%$ respectively for $100\cubicMpc$ and
$1000\cubicMpc$ pointings.

\subsection{Imaging vs. identifying of the counterpart}

Our study focuses on the probability of {\em imaging} the \ac{EM} counterpart to
a detected \ac{GW} signal -- i.e., pointing a telescope so that the \ac{EM}
counterpart is within the field of view -- but not on the probability of
detecting and {\em identifying} it among background sources.  In reality, some
telescopes may have trouble observing weak, distant \ac{EM}
counterparts \cite{LVCOpticalTransients}.

For example, \citet{Metzger2012} suggest that the orphan optical afterglow
expected to accompany a \ac{BNS} merger at $200\Mpc$ will have a peak optical
brightness as faint as $\sim23\unittext{mag}$ when viewed slightly off-axis:
beyond the limiting flux of many telescopes.  Even if they are detected,
contamination from background events may make it difficult to pick out the
correct transient.  Identification of \ac{GW} \ac{EM} counterparts among false
positives is addressed by \citet{Nissanke2013}.  The detectability of \ac{EM} counterparts could be further investigated by considering the capabilities of specific telescopes given the
observing requirements of particular sources (for example, their peak
luminosities, light-curve evolution, etc.).


\subsection{Astrophysical assumptions}

We have made a number of assumptions about the astrophysics underlying \ac{BNS}
merger signals:
\begin{enumerate}
    \item {\it B-band luminosity of the host galaxy -- which traces its star
        formation rate -- is a proxy for the merger rate.}  In fact, if there are long time delays between star formation and binary merger, the total mass of
        the host galaxy, traced by K-band luminosity, might be the more relevant indicator of merger rate.  For example, population synthesis modeling
        suggests that half of all \ac{BNS} mergers may take place in elliptical
        galaxies with little ongoing star formation
        \cite{deFreitasPacheco,OShaughnessy:2010}.  Meanwhile, observational evidence on short gamma ray bursts indicates that about a quarter of them occur in elliptical galaxies \cite{FongBerger:2013}, though selection effects associated with the detection of afterglows that allow the host to be identified could influence this fraction.
    \item {\it The completeness of the galaxy catalog is known precisely.}   In practice, the completeness of the catalog is
        estimated from the expected spatial luminosity density in the local
        Universe ($\sim 0.02\Lten\percubicMpc$ for blue luminosity).  Inaccuracy
        in the estimated completeness may lead to a less-than-optimal ranking of
        tiles on the sky.  We can account for the incompleteness of a catalog by changing the weighting we give to individual galaxies; if the catalog completeness fraction is $\lambda$, then the catalogued luminosity of a given pixel, $L_i$, is multiplied by $\lambda$ when computing the prior, with a prior fraction $1-\lambda$ painted uniformly over the entire \ac{GW} sky uncertainty region to account for the galaxies missed in the catalog.
        \item {\it Mergers are spatially coincident with host galaxies on the celestial
        sphere.}  Natal kicks accompanying supernovae that give birth to the neutron-star components of a binary (up to hundreds of $\unittext{km} \unittext{s}^{-1}$
        \citep{Fryer1997}) can combine to give a significant velocity to the binary as a whole.  As a result, mergers are distributed at larger distances from the galactic center than typical stellar concentrations \cite{FongBerger:2013}, and galaxies should properly be treated as extended objects rather than point sources.  However, for telescope fields of view of order a square degree or more and typical source distances of $100$--$200$ Mpc, treating galaxy sizes $\lesssim 100$ kpc as point sources will not affect our results.  On the other hand, binaries may be completely ejected from their host galaxies  \cite[e.g.,][]{Kelley:2010}, and some fraction of the``no-host'' short
        gamma ray bursts \cite{Berger:2010,Tunnicliffe:2013} may provide evidence for this
        population of merging ejected binaries, which may be separated by more than a Mpc from their host galaxy (but see discussion in \cite{Kanner:2012} and references therein).
\end{enumerate}

We suggest a future study of the importance of these effects -- given our
ignorance -- as parameterized priors.  One would allow nature to choose a {\em
true} value of a given parameter (e.g. the relative contribution of blue and red
luminosity tracers to merger rates) and attempt to image counterparts from the
resulting \ac{GW} events by ranking tiles according to an {\em assumed}
parameter value representing our own knowledge.  The effects of our ignorance of
the true values of each parameter could thus be described by a matrix in which
one dimension represents nature's choice of prior, and the other our assumed
knowledge.

\subsection{Coherent use of galaxy catalogs}

Finally, we have investigated the utility of a galaxy catalog when applied to
the sky location posterior obtained from a parameter estimation pipeline.   In
practice, if a galaxy catalog were to be used for follow-up, it should be
applied as a prior during coherent Bayesian parameter estimation \cite{S6PE}.  Doing so would make it possible to consistently account for the probability that a given galaxy hosts the \ac{GW} source, which depends not only on the galaxy luminosity but also on the distance to the galaxy and the inclination and orientation of the binary, which must yield a \ac{GW} signal amplitude consistent with observations.  This is particularly important when considering the correlations between the recovered \ac{GW} signal parameters such as inclination and distance.   Moreover, using coherent Bayesian parameter estimation would allow complex sky location posteriors could be accurately accounted for.  

%
%



\begin{acknowledgments}
    We thank Edo Berger, Marica Branchesi, Walter Del Pozzo, Will Farr, Jonah Kanner, Mansi Kasliwal, Erik Katsavounidis, Luke Kelley, Drew Keppel, Brian Metzger, Trevor Sidery, and Alberto Vecchio for useful discussions.  CH and IM are grateful to the Kavli Institute for Theoretical
    Physics and the organizers of the ``Chirps, Mergers and Explosions: The
    Final Moments of Coalescing Compact Binaries'' program, supported in part by
    the National Science Foundation under Grant No. NSF PHY11-2591.  They wish
    to thank the participants for many fruitful discussions.  This research was
    supported in part by Perimeter Institute for Theoretical Physics.  Research
    at Perimeter Institute is supported by the Government of Canada through
    Industry Canada and by the Province of Ontario through the Ministry of
    Research and Innovation.
\end{acknowledgments}

\bibliography{catalog}

\end{document}